\documentclass[10pt, conference]{article}
\usepackage[a4paper, left=2cm, right=2cm, top=2.5cm, bottom=2.5cm]{geometry}
\usepackage{graphicx}
\usepackage{amsmath}
\usepackage{amssymb}
\usepackage{verbatim}
\usepackage{multirow}
\usepackage{tikz}
\usetikzlibrary{quantikz2}
\usepackage{float}
\usepackage{subcaption}
\usepackage{array}
\usepackage{pgfplots}
\pgfplotsset{compat=1.18}
\usepackage{amsthm}
\usepackage{url}
\usepackage{hyperref}
\usepackage{xcolor}
\hypersetup{colorlinks=false,
allbordercolors=white}
\usepackage{algorithmic}
\usepackage{algorithm}
\usepackage{authblk}

\usepackage{cite}
\usepackage{amsmath,amssymb,amsfonts}
\usepackage{algorithmic}
\usepackage{graphicx}
\usepackage{textcomp}
\usepackage{xcolor}
 \usepackage{array}
 \usepackage{multirow}
\def\BibTeX{{\rm B\kern-.05em{\sc i\kern-.025em b}\kern-.08em
    T\kern-.1667em\lower.7ex\hbox{E}\kern-.125emX}}

\title{Training Variational Quantum Circuits Using Particle Swarm Optimization\\
}

\author{Marco Mordacci and Michele Amoretti}

\affil{Quantum Software Laboratory, Department of Engineering and Architecture, University of Parma, 43124 Parma, Italy \\
\texttt{marco.mordacci1@unipr.it}}

\date{}

\begin{document}
\maketitle

\begin{abstract}
In this work, the Particle Swarm Optimization (PSO) algorithm has been used to train various Variational Quantum Circuits (VQCs). This approach is motivated by the fact that commonly used gradient-based optimization methods can suffer from the barren plateaus problem.

PSO is a stochastic optimization technique inspired by the collective behavior of a swarm of birds. The dimension of the swarm, the number of iterations of the algorithm, and the number of trainable parameters can be set. In this study, PSO has been used to train the entire structure of VQCs, allowing it to select which quantum gates to apply, the target qubits, and the rotation angle, in case a rotation is chosen. The algorithm is restricted to choosing from four types of gates: Rx, Ry, Rz, and CNOT.

The proposed optimization approach has been tested on various datasets of the MedMNIST, which is a collection of biomedical image datasets designed for image classification tasks.

Performance has been compared with the results achieved by classical stochastic gradient descent applied to a predefined VQC. The results show that the PSO can achieve comparable or even better classification accuracy across multiple datasets, despite the PSO using a lower number of quantum gates than the VQC used with gradient descent optimization.
\end{abstract}

\section{Introduction}
Quantum Machine Learning (QML) leverages quantum systems to tackle complex tasks~\cite{b1}, with Variational Quantum Circuits (VQCs) serving as quantum counterparts of neural networks. In a VQC, rotation angles are trained to perform the desired task, typically via classical gradient descent, but can suffer from trainability issues, such as barren plateaus~\cite{b3}. 

Designing optimal architecture for VQCs is a fundamental task to achieve better results. Quantum Architecture Search (QAS)~\cite{b7} aims to automatically find optimal VQC structures using methods like Reinforcement Learning~\cite{b8}, evolutionary algorithms~\cite{b9}.

In this work, Particle Swarm Optimization (PSO)~\cite{b5}, which was already used in~\cite{b11} to train the parameters of a VQC, has been used to train the entire architecture of the VQC, which means that both rotation angles and the applied gates are chosen by the optimizer.

\section{Particle Swarm Optimization}
PSO is a stochastic optimization technique that simulates the movement of a swarm of birds~\cite{b5}. PSO is characterized by a set of particles and by iterations, which are movements of the swarm in the solution space. 
A particle is characterized by the following parameters: the position in the problem's domain; its velocity; the best position found by the particle; the best position found by the swarm.


Three hyperparameters must be set: $c_1$, $c_2$ and $w$.
At the beginning of the optimization process, the cognitive coefficient $c_1$ is set high and the social coefficient $c_2$ low, encouraging exploration by focusing on each particle's own best solution. As iterations progress, $c_1$ decreases, while $c_2$ increases, shifting the behavior toward exploitation of the swarm's global best. The inertia weight $w$ also starts high to support broad exploration and is gradually reduced to slow particle movement, aiding convergence to an optimal or suboptimal solution.

\section{Results}
In this work, PSO has been used to train the full structure of the VQC. In particular, the algorithm has $50$ particles and has been tested with $40$ and $80$ dimensions, representing trainable parameters. 
The parameters of a particle are grouped in sets of $4$ and define: the applied gate (Rx, Ry, Rz, and CNOT); the target qubit; the control qubit if a CNOT is selected; the rotation angle if a rotation is used.
Thus, with $40$ parameters, the VQC can have at most $10$ gates.
Since the parameters are random numbers in the range $[0, 1]$, this range is discretized:
\begin{equation}
    g = \text{round}(1 + \#gates \cdot parameter\_value)
\end{equation}
where $g$ is the chosen gate. This maps the space $[0, 1]$ to $[1, \#gates]$.
Similarly, for qubit selection, the range $[0, 1]$ is mapped to $[1, \#qubits]$. For rotation angles, the range $[0, 1]$ is scaled to $[0, 2\pi]$.


The PSO process was executed for $100$ iterations and tested on MNIST (digits 0 and 1) and several MedMNIST~\cite{b6} datasets. Only binary tasks were considered by selecting the first two classes: CNV and DME for oct, Bladder and Left Femur for organA, organC, and organS. Other considered datasets were already binary.

PSO has been compared to the Adam optimizer (learning rate of $0.01$, batch size $32$, for $100$ epochs). A VQC with $8$ qubits was used, consisting of two layers of Ry rotations and CNOT gates arranged in a circular topology ($32$ gates and $16$ parameters). Principal component analysis was applied to reduce the input to the $8$ most significant features for angle encoding.

Table~\ref{tab:val_results} shows the accuracy on the validation set, where PSO can achieve similar or better results, except for pneumonia and organS.
Table~\ref{tab:test_results} presents the test results, and a similar behavior can be observed. Furthermore, in the oct dataset, PSO $80$ outperforms Adam, unlike in validation, suggesting a slightly better generalization by the PSO-trained circuit.

\begin{table}[htbp]
    \centering
    \resizebox{0.40\textwidth}{!}{
    \begin{tabular}{|>{\centering\arraybackslash}m{1.5cm}||>{\centering\arraybackslash}m{1.5cm}|>{\centering\arraybackslash}m{1.5cm}|>{\centering\arraybackslash}m{1.5cm}|}
    \hline
         Dataset & Val Adam & Val PSO $40$ & Val PSO $80$ \\
         \hline
         Breast & $76\%$ & $75\%$ & \boldmath$81\%$ \\
         \hline
         Chest & $53\%$ & $52.5\%$ & \boldmath$57\%$ \\
         \hline
         Oct & \boldmath$60\%$ & $56\%$ & $58\%$ \\
         \hline
         Pneumonia & \boldmath$86\%$ & $78\%$ & $78\%$ \\
         \hline
         OrganA & $81.5\%$ & \boldmath$92\%$ & \boldmath$92\%$ \\
         \hline
         OrganC & $85.5\%$ & $85\%$ & \boldmath$86\%$ \\
         \hline
         OrganS & \boldmath$85\%$ & $79\%$ & $81$\% \\
         \hline
    \end{tabular}
    }
    \caption{Accuracies achieved by Adam and PSO with the validation set.}
    \label{tab:val_results}
\end{table}

\vspace{-0.7cm}
\begin{table}[htbp]
    \centering
    \resizebox{0.40\textwidth}{!}{
    \begin{tabular}{|>{\centering\arraybackslash}m{1.5cm}||>{\centering\arraybackslash}m{1.5cm}|>{\centering\arraybackslash}m{1.5cm}|>{\centering\arraybackslash}m{1.5cm}|}
    \hline
         Dataset & Test Adam & Test PSO $40$ & Test PSO $80$ \\
         \hline
         MNIST & $73.7\%$ & \boldmath$74.5\%$ & \boldmath$75\%$ \\
         \hline
         Breast & $71\%$ & \boldmath$75\%$ & \boldmath$73\%$ \\
         \hline
         Chest & $51\%$ & $52\%$ & \boldmath$56\%$ \\
         \hline
         Oct & $54\%$ & $52\%$ & \boldmath$55\%$ \\
         \hline
         Pneumonia & \boldmath$81\%$ & $71\%$ & $72\%$ \\
         \hline
         OrganA & $67\%$ & \boldmath$78\%$ & \boldmath$78\%$ \\
         \hline
         OrganC & \boldmath$75\%$ & $73\%$ & \boldmath$75\%$ \\
         \hline
         OrganS & \boldmath$73\%$ & $72\%$ & $72\%$ \\
         \hline
    \end{tabular}
    }
    \caption{Accuracies achieved by Adam and PSO with the test set.}
    \label{tab:test_results}
\end{table}

Furthermore, Table~\ref{tab:macro_res} reports class-wise metrics on the breast dataset. Although Adam achieved similar test accuracy, it predicts only a single class, indicating that it failed to learn meaningful patterns. In contrast, the two VQCs optimized with PSO have begun to learn both classes.

\begin{table}[htbp]
    \centering
    \resizebox{0.41\textwidth}{!}{
    \begin{tabular}{|>{\centering\arraybackslash}m{1.5cm}||>{\centering\arraybackslash}m{1.3cm}|>{\centering\arraybackslash}m{1.3cm}|>{\centering\arraybackslash}m{1.3cm}|>{\centering\arraybackslash}m{1.3cm}|}
    \hline
         Optimization algorithm & class & Precision & Recall & f1-score \\
         \hline
         \multirow{2}{*}{\centering Adam} & 0 & $0.0$ & $0.0$ & $0.0$ \\
         \cline{2-5}
         & 1 & $0.73$ & $0.99$ & $0.84$ \\
         \hline
         \multirow{2}{*}{\centering PSO $40$} & 0 & $0.67$ & $0.14$ & $0.24$ \\
         \cline{2-5}
         & 1 & $0.76$ & $0.97$ & $0.85$ \\
         \hline
         \multirow{2}{*}{\centering PSO $80$} & 0 & $0.5$ & $0.1$ & $0.16$ \\
         \cline{2-5}
         & 1 & $0.74$ & $0.96$ & $0.84$ \\
         \hline
         
    \end{tabular}
    }
    \caption{Precision, Recall and f1-score achieved in the breast dataset from the different optimization methods.}
    \label{tab:macro_res}
\end{table}

\begin{figure}
    \centering
    \includegraphics[width=0.45\linewidth]{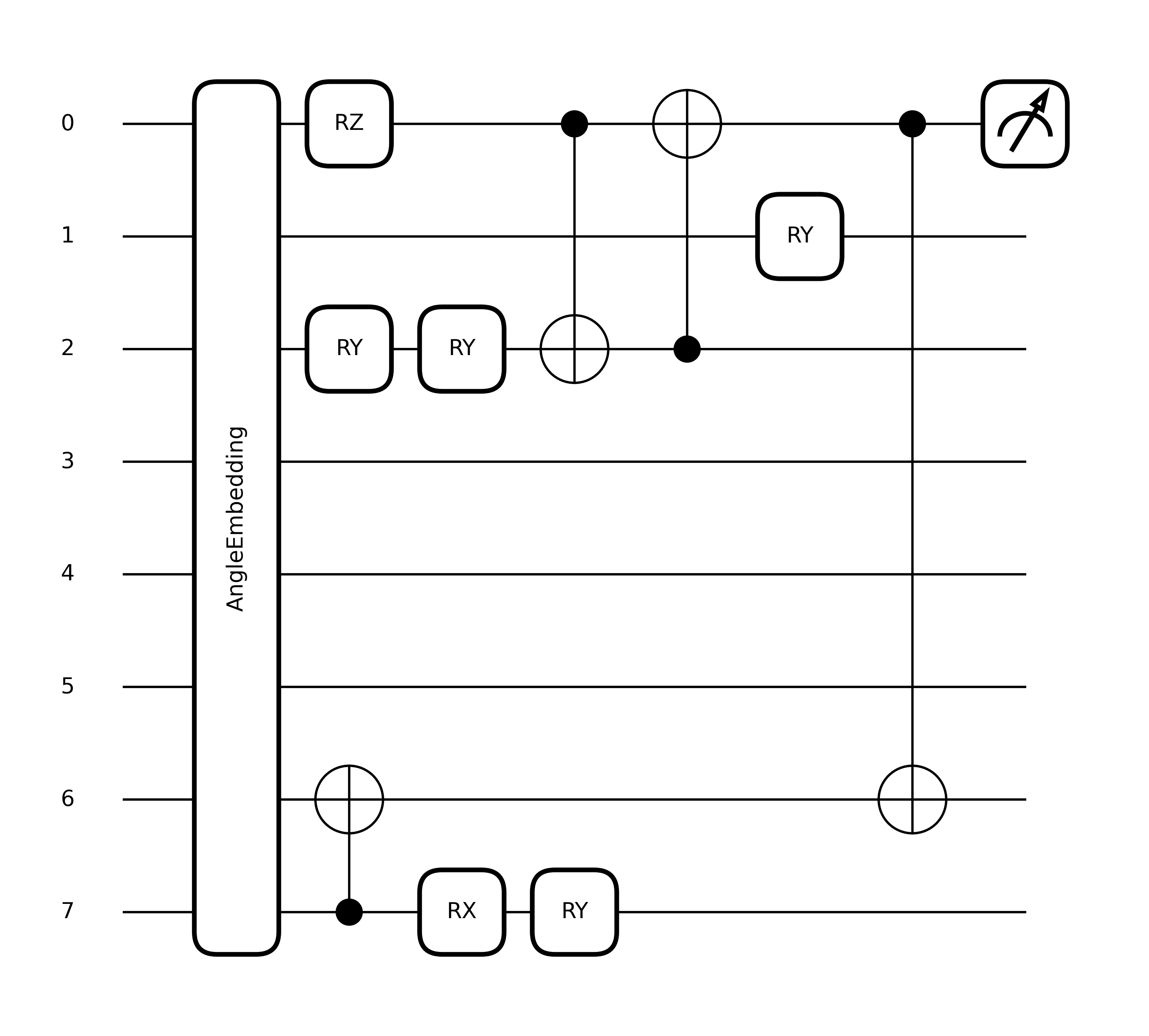}
    \caption{Optimized VQC obtained by PSO during the optimization on the organA dataset, with 40 dimensions. Figure generated using PennyLane.}
    \label{fig:organA40}
\end{figure}

Finally, in Figure~\ref{fig:organA40} an example of VQC achieved by PSO is presented. It is noteworthy that the VQC obtained by PSO achieved similar or even better performance than Adam, while using a lower number of gates (at most $10$ or $20$). Moreover, as shown in Figure~\ref{fig:organA40}, several gates, such as the two rotations on the last qubit, do not contribute to the classification task; thus, the circuit can attain the same results with an even smaller number of effective gates.

Table~\ref{tab:rigetti} presents the results obtained on Rigetti Ankaa-3 and IQM Garnet devices using 1024 shots and a subset of 100 samples from the organA test set. The presence of noise does not impact the performance of the VQC optimized by PSO, which achieves results comparable to those in the ideal (noise-free) scenario.

\begin{table}[htbp]
    \centering
    \resizebox{0.35\textwidth}{!}{
    \begin{tabular}{|>{\centering\arraybackslash}m{1.5cm}||>{\centering\arraybackslash}m{1.5cm}|>{\centering\arraybackslash}m{1.5cm}|>{\centering\arraybackslash}m{1.5cm}|}
    \hline
         Dataset & Simulation & Rigetti Ankaa-3 & IQM garnet\\
         \hline
         organA & $74\%$ & $74\%$ & $74\%$ \\
         \hline
    \end{tabular}
    }
    \caption{Accuracies achieved by PSO $80$ on a subset of the organA testset.}
    \label{tab:rigetti}
\end{table}

\section{Conclusion}
In this work, PSO was used to train the structure of VQCs on MedMNIST datasets. It was evaluated on binary classification tasks, showing better performance than classical gradient descent with fewer gates.

In future work, the training of more complex VQCs will be explored. Additionally, more challenging tasks, such as multiclass classification, will be addressed.

\section*{Acknowledgment}
We acknowledge the financial support from Spoke 10 - ICSC - ``National Research Centre in High Performance Computing, Big Data and Quantum Computing'', funded by European Union – NextGenerationEU.
This research benefits from the High Performance Computing facility of the University of Parma, Italy (HPC.unipr.it).

\bibliographystyle{unsrt}

\end{document}